\documentclass[conference]{IEEEtran}

\ifCLASSOPTIONcompsoc
  \usepackage[nocompress]{cite}
\else
  \usepackage{cite}
\fi

  \usepackage{graphicx}
  \usepackage{array}
  \usepackage{caption}
  \usepackage{subfigure}
  \usepackage{color}
  \usepackage{amsmath,amssymb}

\begin{document}

\title{Performance Evaluation of Secure Multi-party Computation on Heterogeneous Nodes}

\author{Zhou Ni, Rujia Wang \\ Computer Science, Illinois Institute of Technology \\Email: zni1@hawk.iit.edu, rwang67@iit.edu}






\maketitle

\begin{abstract}

Secure multi-party computation (MPC) is a broad cryptographic concept that can be adopted for privacy-preserving computation. With MPC, a number of parties can collaboratively compute a function, without revealing the actual input or output of the plaintext to others. The applications of MPC range from privacy-preserving voting, arithmetic calculation, and large-scale data analysis. From the system perspective, each party in MPC can run on one compute node. The compute nodes of multiple parties could be either homogeneous or heterogeneous; however, the distributed workloads from the MPC protocols tend to be always homogeneous (symmetric).
In this work, we study a representative MPC framework and a set of MPC applications from the system performance perspective. We show the detailed online computation workflow of a state-of-the-art MPC protocol and analyze the root cause of its stall time and performance bottleneck on homogeneous and heterogeneous compute nodes. 
\end{abstract}

\section{Introduction} \label{sec:introduction}
In the era of big data, cross-computing of massive data can provide better support for many domains such as scientific research, healthcare data analysis, and financial analyst. Meanwhile, for many data providers, they cannot share the raw data to the public for information security or benefit reasons. Therefore, privately sharing the data among multiple parties to compute collaboratively is a challenging task that attracts numerous researchers in the cryptographic area to work on.

The high-level solution to compute collaboratively and privately is to adopt secure multi-party computation (MPC).
With complicated building blocks such as Garble Circuit(GC), Oblivious Transfer(OT), or Homomorphic Encryption(HE), it can hide the sensitive information and preserve the privacy between a group of distrusting parties. 
In other words, it can provide data demanders with multi-party collaborative computing capabilities without leaking the original data from data providers.

From the system perspective, MPC has promising research and commercial value, as it expands the scope of traditional distributed computing with the add-on privacy-preserving feature. However, the challenge of implementing MPC on a large-scale distributed system is, the computation and communication overhead coming from the protocol are  \textit{increasing} with the number of participated parties due to the cryptographic protocol's complexity. In contrast, the goal for distributed computing is to distribute the workloads among several compute nodes evenly so that we can exploit the power from parallel computing to \textit{decrease} the overall computation time. Therefore, a more practical setting, \textit{secure two-party computation}(2PC), can be adopted to achieve better performance while providing the data collected from one party is hidden for the other party. Each party can be viewed as a root node that collects data from a number of parties that trust each other. 

The MPC protocol involves interactive computation and communication phases, and even in the two-party setting, if the two parties have imbalanced computation power, the stalls coming from the interactive protocol can be exacerbated. A simple case would be, one party is a laptop while the other party is a powerful server.
Currently, the cryptographic protocol designs do not consider the setting of heterogeneous systems. Therefore, it's not well known how MPC protocols would suffer from such a setting. As a result, in this work, we would like to study the potential performance bottleneck of MPC protocols in an imbalanced system setting and help the system researchers to understand the new set of cryptographic applications. We select ABY \cite{demmler2015aby}, a fast and flexible 2PC framework as our evaluation candidate. We show that on two types of cloud nodes, how imbalanced computation capabilities could harm the online phase performance with several case studies. We further discuss the potential solutions to mitigate such issues.


\section{Background}

\subsection{An overview on MPC}

MPC \cite{Yao82,Yao86,Goldreich87} has significantly advanced the development of privacy-preserving computation with data held by different parties. 
Garbled circuits (GC) \cite{Yao82,Yao86,Beaver:1990:RCS:100216.100287} were first proposed for MPC: creating the garbled circuit and then securely evaluating the result of a function without disclosing private inputs by different parties. To date, with appropriate composition \cite{Goldreich,Lindell03Compose,LindellBook}, MPC framework can be designed using a wide variety of cryptographic primitives such as garbled circuit (GC) \cite{Beaver:1990:RCS:100216.100287}, homomorphic encryption (HE) \cite{Paillier99,Gentry:2009:FHE:1834954}, secret sharing (SS) \cite{Shamir:1979:SS:359168.359176} and oblivious transfer(OT) \cite{rabin2005exchange}.

The core of the mainstream MPC frameworks primarily use GC, OT and SS as their building block. 
For example, in Yao's GC, the public function is written as a GC; the sender will ``garble" the circuit based on its own inputs, and send the ``garbled" circuit to the receiver. Then, both parties apply OT to generate the receiver's ``garbled" inputs. The receiver lastly evaluates the circuit using all ``garbled" inputs. The communication rounds for Yao's GC is constant but requires symmetric cryptographic operations in the online phase, which could be slow.
In contrast, another set of frameworks, GMW \cite{goldreich2019play} and BGW \cite{asharov2017full}, utilize SS to split the secret into multiple shares. Each party secret-shares their input, and then perform entire computation and GC evaluation on the shares instead of the whole inputs. The schemes used to split secrets can either use Boolean circuits (XOR, AND, MUX) or Arithmetic circuits(addition, multiplication).
With SS, each party can pre-compute all cryptographic operations and reduce the online phase computation time.

\subsection{Implementing MPC on heterogeneous systems}
MPC can be widely used to privately calculate a number of applications, such as the millionaire problem \cite{ioannidis2003efficient}, private set intersection \cite{hazay2017scalable}, inner-product \cite{launchbury2014application} and so on. Ideally, any computing system can be as one party during the MPC process. A more realistic case is that the parties will not have the exactly same computation power. However, current MPC protocols are mostly symmetric in terms of communication and computation overhead. 

While deploying the MPC protocols, the heterogeneity of the computational capabilities may result in poor resource utilization. For example, a weak party with limited computation power may delay the overall MPC computation if we do not consider such constraints in the MPC protocol design. In another word, the workload on each party are not proportional to the computation power.

To understand how MPC will perform on homogeneous and heterogeneous nodes, in this work, we evaluate selected applications and different underlying secret sharing methods with a 2PC framework, ABY \cite{demmler2015aby}. The goal of our work is to show how much under-utilization could happen with heterogeneous party setting versus homogeneous party setting, and envision the approaches to mitigate such resource under-utilization. 


\section{Case study: ABY Framework}

\subsection{Supported Sharing Schemes}
ABY is a novel 2PC framework that supports mixed MPC protocols in the semi-honest adversary model. In ABY, it supports three types of secret sharing: Arithmetic, Boolean, and Yao's GC. 
In this paper, we focus on the Arithmetic sharing and Boolean sharing which are proved have a good online phase efficiency with low latency during communication. The Arithmetic sharing, which was first studied in \cite{cascudo2011torsion}, is based on additive sharing private values between parties. The Arithmetic sharing semantics is constructed by three parts: shared values, sharing, and reconstruction. The shared variable x is denoted as $\langle x\rangle^A$ and the individual share of $\langle x\rangle^A$ is held by party $P_i$ as $\langle x\rangle^{A}_i$. First, we have $\langle x\rangle^{A}_0+\langle x\rangle^{A}_1\equiv x$ (mod $2^l$) with $\langle x\rangle^{A}_0, \langle x\rangle^{A}_1 \in \mathbb{Z}_{2^l}$ from an $l$-bit Arithmetic Sharing $\langle x\rangle^A$ of $x$. Next, in the sharing part, $P_i$ chooses $ r \in_{R} \mathbb{Z}_{2^l}$, sets $\langle x\rangle^{A}_i = x-r$, and sends $r$ to $P_{1-i}$, who sets $\langle x\rangle^{A}_{1-i} = r$. Finally, for reconstruction, $P_{i-1}$ sends its share $\langle x\rangle^{A}_{1-i}$ to $P_i$ who computes $x = \langle x\rangle^{A}_0 + \langle x\rangle^{A}_1$. The $P_i$ obtains the value of $x$ as output is denoted as $x = Rec^{A}_{i}(\langle x\rangle^A)$ at reconstruction process. After both parties obtain the value $x$, then we have $Rec^{A}(\langle x\rangle^A)$. 

As for the Boolean sharing, it uses an XOR-based secret sharing scheme to share a variable. The Boolean sharing also has three parts which are defined by its sharing semantics. A Boolean share $\langle x\rangle^B$ of a bit $x$ is first shared between parties:  $\langle x\rangle^{B}_0 \oplus \langle x\rangle^{B}_1 = x$ with $\langle x\rangle^{B}_0, \langle x\rangle^{B}_1 \in \mathbb{Z}_{2}$ . Then, $P_i$ chooses $ r \in_{R} \{0,1\} $ and computes $\langle x\rangle^{B}_i = x \oplus{r}$. $P_{1-i}$ receives $r$ from $P_i$ and sets $\langle x\rangle^{B}_{1-i} = r$. After sharing, $P_{i-1}$ sends its share $\langle x\rangle^{B}_{1-i}$ to $P_i$ who computes $x = \langle x\rangle^{B}_0 \oplus{\langle x\rangle^{B}_1}$. Although Boolean and Arithmetic sharing have a similar sharing semantics, the construction process of these two types of garble circuits is different. As mentioned in Section 2, the Arithmetic circuit is a sequence of addition and multiplication gates, while the Boolean sharing consist of XOR and AND gates. Despite all that, both of the sharing schemes can work with the same steps in the framework.

\subsection{Breakdown the Online Phase}
\begin{figure}[h!]
\centering
\includegraphics[width = 0.48\textwidth]{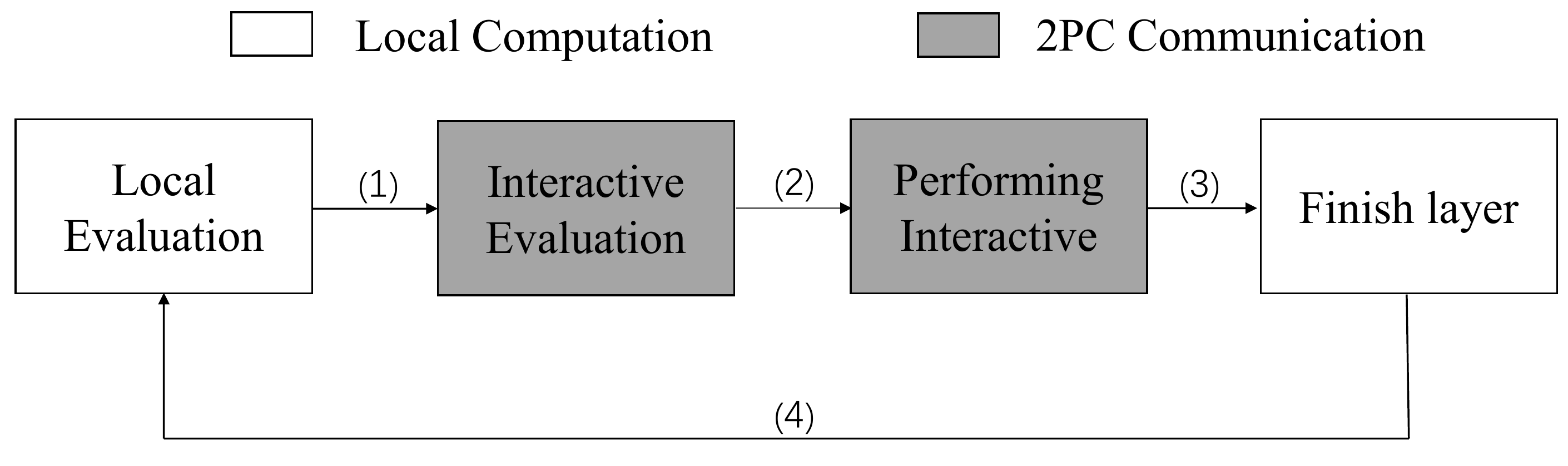}
\caption{ABY online phase}
\label{fig:aby}
\end{figure}

The entire execution of ABY can be divided into the offline phase and the online phase. During the offline phase, the framework starts with preprocessing, such as system initialization, setup, and GC generation. Once the offline phase is done, the framework starts the online phase that contains several iterations of computation and communication. As shown in Figure \ref{fig:aby}, there are four steps in the online phase: local evaluation, interactive evaluation, performing interactive, and finish layer evaluation.
After the pre-computation at the offline phase, the addition of Arithmetic sharing and XOR of Boolean sharing operations are computed at the local evaluation layer. This step is computation only; therefore, a faster machine can complete the task with less computation time. 

Then, shared values are sent for the interactive evaluation.
For example, in the arithmetic sharing, the arithmetic circuit consists of addition and multiplication gates. Therefore, in the interactive evaluation, each party needs first to find out all the arithmetic gates and then distinguish the type of each gate. More specifically, party A has to find out which gates are its own input gates and which gates belong to party B. As for the output gates, party A also has to figure out whether the gates belong to itself or party B's or both of them. After this interactive evaluation, the two parties can perform interactive communications so that they can share their value and then reconstruct the value they shared privately. The time of the value sharing and reconstruction is considered as communication time, which consists of the time stalled when waiting for data sharing from the other party. 
The last step in the online phase is the finish layer evaluation, where the shared values are computed at the local machine again after each party reconstructs their receive value. 

After all these three evaluation processes, a whole single round of online phase is completed, and the next round starts from the first step. The whole online phase matches the sharing semantics, which we defined before.

Clearly, the online phase of the ABY has computation-intensive and communication-intensive steps, and the interactive operations between parties highly rely on the shared secret data received from the other party. In other words, there is a synchronization barrier between the steps in the online phase. When the parties are running on two machines with similar system specification, the data sharing can finish almost simultaneously. However, if the two parties are on two heterogeneous machines, the synchronizations have to be delayed due to the slower party in this case. In the next section, we show an in-depth performance analysis on the online phase and study how the input size, sharing scheme, application, system setting could impact the overall online phase time. 


\section{Evaluated systems and applications}

\subsection{System Setup}
We deploy the ABY framework on two compute nodes on the Chameleon cloud.
We select two types of instances: Low Power Xeon and Atom. The hardware configurations are shown in Table \ref{tab:sys}. The Atom node is considered to be a weak node, while the Low Power Xeon is a strong node. In the experiments, we let the two parties run both on Xeon, both on Atom, and one on each. The results of the experiments are the average of 10 executions unless stated otherwise.



\begin{table}[h]
\centering
    \caption{System setup}
    \label{tab:sys}
\begin{tabular}{|l||c|c|}
\hline
Node Type                                                       & Low Power Xeon                                                        & Atom                                                            \\ \hline
CPU version                                                     & \begin{tabular}[c]{@{}c@{}}Intel E3-1284L v4 \\ @2.90GHz\end{tabular} & \begin{tabular}[c]{@{}c@{}}Intel C2750 \\ @2.40GHz\end{tabular} \\ \hline
Cores/Threads                                                   & 4/8                                                                   & 8/8                                                             \\ \hline
\begin{tabular}[c]{@{}l@{}}Cache\\ (L1d/L1i/L2/L3)\end{tabular} & \begin{tabular}[c]{@{}c@{}}32KB/32KB/\\ 256KB/6MB\end{tabular}        & \begin{tabular}[c]{@{}c@{}}24KB/32KB/\\ 1MB/X\end{tabular}      \\ \hline
RAM Size (GB)                                                   & 32                                                                    & 32                                                              \\ \hline
\end{tabular}
\end{table}



\subsection{Dataset and Application}
We evaluate two privacy-preserving applications in our experiments. The first one is to compute the \textit{inner product} of two vectors using arithmetic sharing. Each party holds only one input vector, and both parties would like to receive the output of the computation. The vector size is considered as the input size that we can adjust for benchmarking. Each element in the vector is set at 16-bit long.

The second application is the \textit{millionaire probability} problem, which is a classical application used in MPC. Boolean sharing is used to compute the results. Random values are given to the two parties as to the input, and the application finds out which party holds more significant value and send the answer to both parties. The bit length of the input value can be adjusted for benchmarking.
The symmetric key length is set at 128 in all cases. 


\section{Results and Analysis}
\subsection{Inner Product}
In this section, we show the execution time breakdown of the online phase with the inner product application. The experiments are done on three sets of systems with small and large input sizes. We set the number of elements in the input vector to $2^{7}$ as the small input size and $2^{17}$ as the large input size.

In each table, we report the time spent on each step of the online phase. The local gates, interactive gate, and layer finish denote the communication time spent on each node. The communication time denotes the time spent waiting for the secret data from the other party. The online phase time sum up all the time spent on previous steps.


\begin{table}[h]
    \centering
    \caption{Inner Product with Atom Nodes}
    \label{tab:inner_atom}    
    \begin{tabular}{|c|c|c|c|c|}
    \hline
       & \multicolumn{2}{c|}{Small} & \multicolumn{2}{c|}{Large} \\ 
    \hline
    Arithmetic local gates(ms) & 0.054 & 0.045 & 47.570 & 47.528 \\ 
    \hline
    Interactive gate(ms) &  0.048 & 0.042 & 17.508 & 17.529 \\ 
    \hline
    Layer finish(ms) & 0.060 & 0.060 & 56.929 & 56.046 \\ 
    \hline
    \textbf{Communication(ms)} & 0.823 & 0.708 & 3.882 & 3.652  \\ 
     \hline
    Online phase(ms) &  1.563 & 1.526 & 154.869 & 154.853\\ 
    \hline
    \end{tabular}
\end{table}


\begin{table}[h]
    \centering
    \caption{Inner Product with Xeon Nodes}
    \label{tab:inner_xeon}  
     \begin{tabular}{|c|c|c|c|c|}
    \hline
       & \multicolumn{2}{c|}{Small} & \multicolumn{2}{c|}{Large} \\ 
    \hline
    Arithmetic local gates(ms) & 0.015 & 0.030 & 14.685 & 14.782 \\ 
    \hline
    Interactive gate(ms) & 0.020 & 0.023 & 7.323 & 7.212 \\ 
    \hline
    Layer finish(ms) & 0.024 & 0.034 & 19.472 & 19.355\\ 
    \hline
    \textbf{Communication(ms)}& 0.218 & 0.172& 1.035 & 1.002  \\ 
     \hline
    Online phase(ms) & 0.474 & 0.479 & 52.234 & 52.098\\ 
    \hline
    \end{tabular}
\end{table}
Table \ref{tab:inner_atom} and \ref{tab:inner_xeon} show the experimental results on two homogeneous nodes setting. We can observe that, with small input size, on two types of system, the communication time dominates the overall online phase time. When the input size increases, on both systems, the major online time is spent on computation. Since the two machines in each system have the same computation power, the overall stall time due to the computation is minimal(around 2\%) with large input size.


\begin{table}[h]
    \centering
    \caption{Inner Product on Heterogeneous Nodes}
    \label{tab:inner_hete}    
    \begin{tabular}{|c|c|c|}
    \hline
    Small input & Xeon & Atom \\
    \hline
    Arithmetic local gates(ms) & 0.024 & 0.048\\
    \hline
    Interactive gate(ms) & 0.024 & 0.053\\
    \hline
    Layer finish(ms) & 0.032 & 0.059\\
    \hline
    \textbf{Communication(ms)} & 1.016 & 0.489\\
    \hline
    Online phase(ms) & 1.434 & 1.316\\
    \hline
    \hline
    Large input & Xeon & Atom \\
    \hline
    Arithmetic local gates(ms) & 14.339 & 47.341\\
    \hline
    Interactive gate(ms) & 7.545 & 17.598\\
    \hline
    Layer finish(ms) & 29.688 & 57.397\\
    \hline
    \textbf{Communication(ms)}& 98.449 & 1.812\\
    \hline 
    Online phase(ms) & 156.724 & 156.712\\
    \hline
    \end{tabular}
\end{table}

Next, we deploy the framework on a heterogeneous system so that the two parties have different computation capability. As shown in Table \ref{tab:inner_hete}, in this setting, we are observing a skewed communication time between the two parties.
The party on the Xeon machine always has a longer communication time compared with the party on Atom machine: 2.07x with a small input, and 54.33x with a large input. The reason is that the workload assigned to both parties is the same; however, the Atom node processes the computation much slower compared to the Xeon node, especially when the input size is growing. With the large input, the majority time spent on Xeon is due to the stall: the communication time is 62.8\% of total online time. Clearly, such a situation is aggravated if we want to compute more complicated tasks or a larger scale of inputs.


\subsection{Millionaire Probability}

Similarly, we show our experimental results for the millionaire probability application. We vary the input bit length to show the results for small ($2^{5}$) and large ($2^{15}$) input size.
The results of performance on homogeneous nodes are shown in Table \ref{tab:mill_atom} and \ref{tab:mill_xeon}. We observe the same performance trends as the inner product application: when the input size is small, the communication time dominates (over 50\%); while the input size is large, more time is spent on computation, only around 5\% time is spent on the communication. This observation confirms that small-scale 2PC is bounded by the communication link, while large-scale 2PC is still bounded by the computation. Besides, the boolean gates tend to add less pressure on the local gate evaluation stage, compared with the arithmetic gates.



\begin{table}[ht]
    \centering
    \caption{Millionaire Probability with Atom Nodes}
    \label{tab:mill_atom}
     \begin{tabular}{|c|c|c|c|c|}
  
    \hline
       & \multicolumn{2}{c|}{Small} & \multicolumn{2}{c|}{Large} \\ 
    \hline
    Boolean local gates(ms) & 0.032 & 0.035 & 26.820 & 26.723 \\ 
    \hline
    Interactive gate(ms) &  0.079 & 0.086 & 31.965 & 31.971 \\ 
    \hline
    Layer finish(ms) & 0.032 & 0.031 & 16.038 & 16.613 \\ 
    \hline
    \textbf{Communication(ms)} & 2.249 & 2.216 & 5.944 & 5.316  \\ 
     \hline
    Online phase(ms) &  3.482 & 3.575 & 83.331 & 83.168\\ 
    \hline
    \end{tabular}
\end{table}

\begin{table}[ht]
    \centering
        \caption{Millionaire Probability with Xeon Nodes}
         \label{tab:mill_xeon}
     \begin{tabular}{|c|c|c|c|c|}
      
    \hline
       & \multicolumn{2}{c|}{Small} & \multicolumn{2}{c|}{Large} \\ 
    \hline
    Boolean local gates(ms) & 0.011 & 0.020 & 7.073 & 7.127 \\ 
    \hline
    Interactive gate(ms) &  0.028 & 0.034 & 9.318 & 9.588 \\ 
    \hline
    Layer finish(ms) & 0.010 & 0.017 & 5.365 & 5.561 \\ 
    \hline
    \textbf{Communication(ms)} & 0.453 & 0.419 & 1.534 & 1.066  \\ 
     \hline
    Online phase(ms) &  0.847 & 0.865 & 24.014 & 24.066\\ 
    \hline
    \end{tabular}
\end{table}

Table \ref{tab:mill_hete} shows how the heterogeneous system would have an impact on the communication time on both nodes. We observe the imbalanced computation time on both nodes, similar to the case study on the inner product application above. For the small input, the communication time on Xeon is 1.7x over Atom; for the large input, the communication time on Xeon is significantly higher, which is 15.2x over Atom. Obviously, the overall online phase performance is limited by the slower party in the system. The stall time caused by imbalanced computation power on Xeon with a large input is 69\%.

\begin{table}[ht]
    \centering
    \caption{Millionaire Probability on Heterogeneous Nodes}
    \label{tab:mill_hete}
    \begin{tabular}{|c|c|c|}
    \hline
    Small input & Xeon & Atom \\
    \hline

    Boolean local gates(ms) & 0.019 & 0.028\\
    \hline
    Interactive gate(ms) & 0.043 & 0.071\\
    \hline
    Layer finish(ms) & 0.022 & 0.029\\
    \hline
    \textbf{Communication(ms)} & 1.184 & 0.679\\
    \hline
    Online phase(ms) & 1.718 & 1.670\\
    \hline
    \hline
    Large input & Xeon & Atom \\
    \hline
    Boolean local gates(ms) & 7.594 & 26.688\\
    \hline
    Interactive gate(ms) & 9.106 & 31.446\\
    \hline
    Layer finish(ms) & 6.81 & 15.337\\
    \hline
    \textbf{Communication(ms)}& 54.726 & 3.608\\
    \hline
    Online phase(ms) & 79.323 & 79.398\\
    \hline
    \end{tabular}
\end{table}
\subsection{Sensitivity Study}
\begin{figure}[ht]
\centering
\includegraphics[width = 0.45\textwidth]{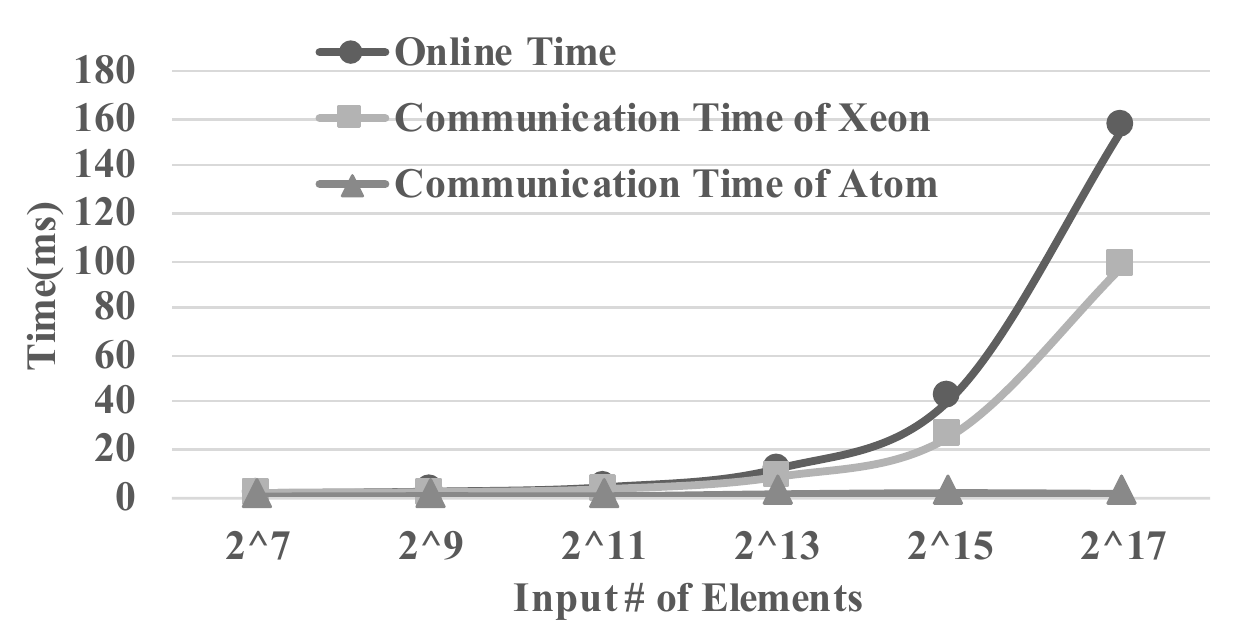}
\caption{Inner Product}
\label{fig:sen-inner}
\end{figure}

\begin{figure}[ht]
\centering
\includegraphics[width = 0.45\textwidth]{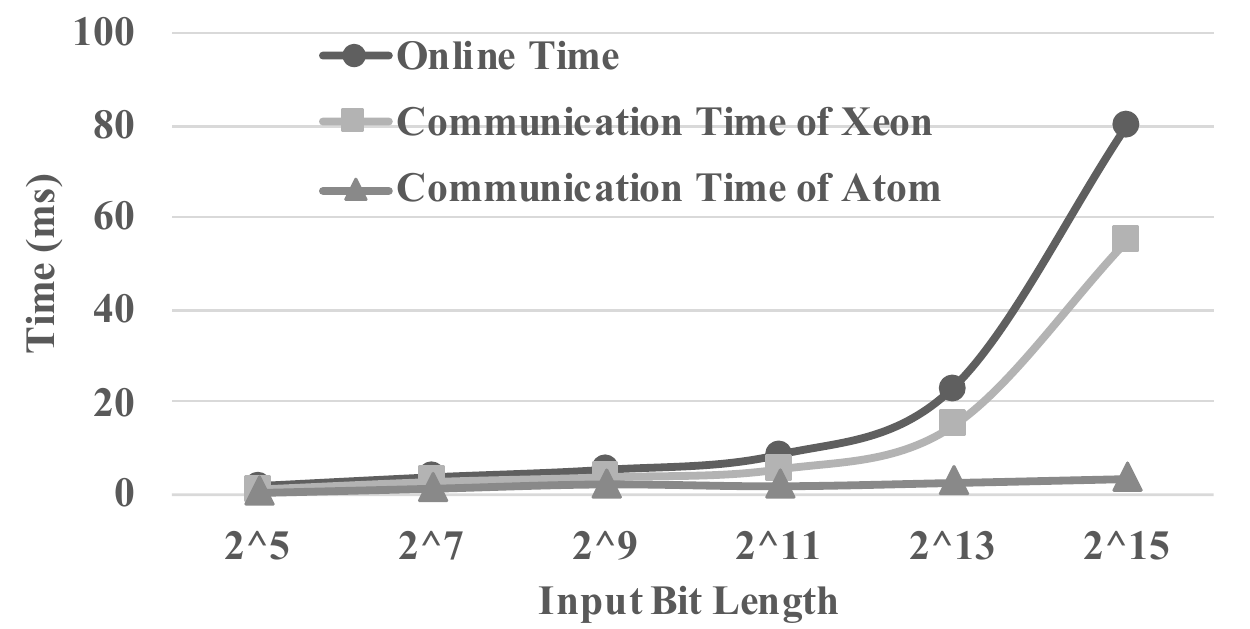}
\caption{Millionaire Probability}
\label{fig:sen-mill}
\end{figure}

Lastly, we show how the input size can impact the overall online time and the communication time on Xeon/Atom nodes in a heterogeneous system. Figure \ref{fig:sen-inner} and \ref{fig:sen-mill} show the trend of the reported metrics with various input sizes. With the increasing input size, the Xeon node has an exponential increasing communication time due to longer computation on the Atom node. The root cause for such an imbalanced stall time increase is because the initial 2PC workload distribution does not consider the underlying computing node's characteristics. Such imbalance is aggravated when the computation load is increasing. Based on the characterization, a more reasonable MPC framework should consider the efficiency in not only terms of cryptography complexity, but also the underlying system that executes the protocols.


\section{Conclusions and Future Work}
In this work, we start a system-level performance analysis on a selected MPC framework, ABY, to show the root cause of the performance bound on real systems. We identify the common workflow for the secret data sharing and identify that in a heterogeneous system with biased computation power, the current protocol could cause an increasing stall time with larger input size.

This work is the first step to analyze and optimize the MPC protocols on systems.
Our future work includes: 1) characterize MPC protocol with increasing party number on the system-level; 2) accelerate the MPC applications through system-level optimization, such as overlapping the computation with communication; 3) co-designed MPC protocols to enable reasonable data sharing with the consideration of underlying party's computation capability. The ultimate design goal is to combine the system and cryptographic optimization approaches to minimize the computation stall time due to the imbalanced nodes.





\bibliographystyle{plain}
\bibliography{rw.bib,ref.bib}

\end{document}